# Strong impact of low-level substitution of Mn by Fe on the magnetoelectric coupling in $TbMnO_3$

A. Maia[1], R. Vilarinho[2], P. Proschek[3], M. Lebeda[1], M. Mihalik jr.[4], M. Mihalik[4], P. Manuel[5], D. D. Khalyavin[5], S. Kamba[1], J. Agostinho Moreira[2]

[1]*Institute of Physics of the Czech Academy of Sciences, Na Slovance 2, 182 00 Prague, Czech Republic*

[2]*IFIMUP, Physics and Astronomy Department, Faculty of Sciences, University of Porto, Porto, Portugal*

[3]*Faculty of Mathematics and Physics, Charles University, Ke Karlovu 5, 121 16 Prague, Czech Republic*

[4]*Institute of Experimental Physics Slovak Academy of Sciences, Watsonova 47, Košice, Slovak Republic*
[5]*ISIS Facility, Rutherford Appleton Laboratory, Harwell Campus, Didcot OX11 0QX, UK.*

## Abstract

The correlation between static magnetoelectric coupling and magnetic structure was investigated in $TbMn_{0.98}Fe_{0.02}O_3$ with magnetic field up to 8 T and down to 2 K. Single crystal neutron diffraction experiments reveal a substantial increase in the temperature dependence of the incommensurate modulation wavevector of the antiferromagnetic phase, as the magnetic field strength increases. Magnetic field dependent pyroelectric current measurements revealed significantly higher magnetoelectric coupling at magnetic fields below 4 T than in pure $TbMnO_3$. This is due to the higher sensitivity of the incommensurably modulated cycloid structure to weak magnetic fields. Detailed analysis of our data confirmed that the ferroelectric polarization is induced by inverse Dzyaloshinskii–Moriya interaction for magnetic field strength up to 4 T, but at higher fields a departure from theoretical predictions is ascertained, giving evidence for an additional, as yet misunderstood, contribution to magnetoelectric coupling. It shows that a small 2% substitution of $Mn^{3+}$ by $Fe^{3+}$ has strong impact on the magnetic structure, promoting the destabilization of the incommensurably modulated magnetic cycloidal structure of $TbMnO_3$ in a magnetic field above 5 T. We demonstrate that the magnetoelectric coupling magnitude can be tuned through suitable substitutional elements, even at low level, inducing local lattice distortions with different electronic and magnetic properties.

## 1. Introduction

Magnetoelectric multiferroics, where spontaneous long-range magnetic ordering and ferroelectricity coexist and are coupled, represent a very attractive class of compounds combining rich and fascinating fundamental physics [1,2]. A key point to be studied in this report is the effect of controlled chemical substitution on the relationship between structure, magnetism, ferroelectricity and magnetoelectric coupling. In this regard, the solid solution $TbMn_{1-x}Fe_xO_3$ proved to be a suitable system to unravel such delicate coupling between magnetism and ferroelectricity [3–6]. Although the $Mn^{3+}$ cation is Jahn-Teller active, unlike the $Fe^{3+}$ cation, they have the same ionic radius for the $6^{th}$ coordination, allowing to synthetize isostructural solid solutions in all $x$-range. However, the substitution of $Mn^{3+}$ by $Fe^{3+}$ implies a change in the octahedra distortions of perovskite lattice, with high impact on the magnetic and polar properties. Previous experimental studies in $TbMn_{1-x}Fe_xO_3$ at ambient temperature have evidence for the linear decrease of the cooperative Jahn-Teller distortion amplitude for $0 \leq x \leq 0.5$, and its suppression for higher Fe-concentrations [7]. Furthermore, the octahedra tilting has been found to be sensitive to the amplitude of the Jahn-Teller distortion, which gives clear evidence for the role played by the interplay between structure and magnetism, with impact on the ferroelectric stabilization [7].

$TbMnO_3$ is a type-II multiferroic [8] prototype with a large magnetoelectric coupling. Its phase diagram can be summarized as follows. At room conditions, $TbMnO_3$ is a paraelectric paramagnet with orthorhombic crystal structure described by the *Pbnm* space group [8]. At $T_N = 41$ K, $TbMnO_3$ undergoes an antiferromagnetic (AFM) phase transition, with incommensurate sinusoidal collinear order of the $Mn^{3+}$ spins [9,10]. In this phase, the spins are aligned along the *b*-axis with a temperature dependent propagation vector, taking the value $\boldsymbol{q}_m = (0\ 0.29\ 1)$ at $T_N$ [9,10]. Accompanying the magnetic modulation structure, there is a lattice modulation with wavevector found to be $\boldsymbol{k}_l = 2\boldsymbol{q}_m$ [11,12]. Below $T_C = 28$ K, the sinusoidal spin modulation turns into a cycloidal spiral in the *bc*-plane, as revealed by neutron diffraction studies [9,10]. Whether the magnetic modulation remains incommensurate or changes into commensurate is still a matter of debate, since it becomes weakly temperature dependent below $T_C$ but not completely locking in a fixed commensurate value [9–11,13]. The emergence of a spontaneous electric polarization along the *c*-axis coincides with the cycloid spin ordering below $T_C$, implying an intrinsic entanglement between the spiral-type antiferromagnetism and ferroelectricity [8,14]. According to N. Aliouane *et al* [15], $TbMnO_3$ exhibits quadratic magnetoelastic coupling in the absence of applied magnetic field. Still, a linear magnetoelastic coupling emerges in $TbMnO_3$ when an external magnetic field is applied along the *b*-axis, so that ferroelectricity is no longer a secondary effect [15]. The linear magnetoelastic coupling drives a magneto-structural transition from the cycloidal modulated phase, with the spontaneous polarization along the *c*-axis, to a phase with the spontaneous polarization along the *a*-axis, for a magnetic field higher than 4.5 T [14]. When the external field is applied along the *a*-axis, a similar flop is found but at higher critical field of about 9 T [14]. Upon further temperature decrease, a quasi-long-range ordering of the *f* electrons spins of the $Tb^{3+}$ ions takes place below $T_N^{Tb} \approx 7$ K [9,10].

In general, the isovalent partial substitution of Mn by Fe ($TbMn_{1-x}Fe_xO_3$) modifies the sequence of phase transitions [5]. However, for low-level substitution, corresponding to $0 < x < 0.05$, the phase sequence is found to be the same as in $TbMnO_3$, although both $T_N$ and $T_C$ decrease with increasing *x*, while $T_N^{Tb}$ barely shifts [6]. The critical temperatures decrease with increasing *x*, gives clear evidence for the destabilization of the magnetic interactions underlying the magnetic properties [6]. The main outcome concerns the decrease of the maximum electric polarization as $Fe^{3+}$ content increases towards $x = 0.04$, which has been attributed to the gradual fading out of the cycloidal spin ordering with increasing *x*, and the suppression of ferroelectricity for $x \geq 0.05$ [6].

Most of the previous studies in the multiferroic region of the $TbMn_{1-x}Fe_xO_3$ solid solution ($0 < x < 0.05$) relied on polycrystalline samples, losing information regarding the anisotropic properties of both polarization and magnetism [4,6]. Because such features play important roles in the pure compound, they are assumed to be particularly important in the multiferroic region of the solid solution as well. In particular, the direction of the applied magnetic field should have interesting effects on the magnetic structure, some of which inferred by the direction of polarization switching from the *c*- to the *a*-axis [14]. To understand in more detail the magnetic structure and magnetoelectric coupling of low-level Fe-substituted $TbMnO_3$, we performed a detailed study of the temperature and magnetic field dependence of the ferroelectric polarization of oriented $TbMn_{0.98}Fe_{0.02}O_3$ single crystals. Neutron diffraction data in $TbMn_{0.98}Fe_{0.02}O_3$ are discussed in connection with the results of polarization measurements, towards the characterization of the magnetoelectric coupling. The systematic study of magnetoelectric phenomena provides a deeper understanding of the effect of the B-site cation on the microscopic mechanisms underlying the magnetic and polar properties in $TbMnO_3$.

## 2. Experimental Details

High quality $TbMn_{0.98}Fe_{0.02}O_3$ single crystals were grown by floating zone method in an FZ-T-4000 (Crystal Systems Corporation) mirror furnace, using $MnO_2$, $Tb_4O_7$ and $Fe_2O_3$ as starting materials. The starting powders were mixed in the intended Tb:Mn:Fe stoichiometric ratio, cold pressed into rods and sintered at 1100 °C for 12–14 h in air. Growing was performed in air atmosphere, feed and seed rod were rotated with a speed of 30 rpm in opposite directions and a pulling speed of 6 mm/h was used. The typical length of a grown ingot was between 3 and 5 cm [3,4]. The obtained ingot was oriented using Laue diffraction patterns and special care was taken to avoid macle. The samples were cut from a macle free region of the oriented ingot. The samples were first characterized by X-ray diffraction and Raman scattering at room temperature, and temperature dependent magnetization and specific heat. The obtained results are in good agreement with published data [3,4].

The pyroelectric currents were measured in a standard Quantum Design PPMS while heating at a rate of 5 $Kmin^{-1}$, after cooling under a fixed applied magnetic field. The pyroelectric current measurements were performed after poling the crystals with an electric field of 100 $Vmm^{-1}$, while cooling from a temperature above the Néel temperature. Before measuring, the poling electric field was turned off at 2 K. The temperature dependence of the electric polarization was obtained through time integration of the measured pyroelectric currents.

Single crystal neutron diffraction under applied magnetic field was measured at Wish beamline of the ISIS Neutron and Muon Source at the Rutherford Appleton Laboratory of the Science and Technology Facilities Council. Single crystal neutron diffraction patterns were recorded at fixed temperatures between 5 and 50 K after cooling the sample under an applied magnetic field along the *b*-axis, up to 8 T. The sample was inserted in a helium flow Oxford cryostat, placed in such a way that the *b*-axis is oriented in the vertical direction, parallel to the applied magnetic field, generated by a 9 T magnet. This magnet has an 80° window in the horizontal and a 30° one in the vertical directions. Due to the magnet surrounding the sample, only a restricted angle range could be measured. With this configuration, limited number of nuclear and magnetic diffraction peaks could be studied as a function of temperature and magnetic field, with the condition that the (*hkl*) peaks have $k = 0$ or $k = q_{Mn}$.

## 3. Results

### 3.1. Polar properties under magnetic field

This section is focused on the temperature dependence of the electric polarization, obtained in the following measurement configurations: ***E*** || *a* and ***B*** || *b*, and ***E*** || *c* and ***B*** || *a*, *b*, and *c*, for which the largest effects of the applied magnetic field on the polar properties of $TbMn_{0.98}Fe_{0.02}O_3$ are found.

Figure 1 (a) and Figure 1 (b) show the temperature dependence of the pyroelectric current density, $J(T)$, of $TbMn_{0.98}Fe_{0.02}O_3$ measured along the *a*- and *c*-axes, respectively, under an applied magnetic field (0 – 9 T) along the *b*-axis. The corresponding temperature dependence of the electric polarization, along the *a*- and *c*-axis, obtained from the time integration of the pyroelectric current density, is presented in Figure 1 (c) and Figure 1 (d), respectively. At 0 T, the pyroelectric current density measured along the *c*-axis, $J_c(T)$, peaks at $T_C$ = 23 K, as expected according to earlier reports for $TbMn_{0.98}Fe_{0.02}O_3$ ceramics, evidencing the onset of the ferroelectric phase, with spontaneous electric polarization along the *c*-axis [6]. This anomaly slightly downshifts with increasing magnetic field strength, reaching 21 K at 9 T.

A second anomaly in $J_c(T)$ is observed at lower temperatures, but only for a magnetic field strength above 2 T. This anomaly peaks in opposite sense relatively to the first one, and monotonously upshifts from 6 K, for 2 T, to 14 K, for 9 T. Its amplitude exhibits a non-monotonous magnetic field dependence, being maximum at 5 T. At 0 T, $P_c(T)$ monotonously increases on cooling from $T_C$ = 23 K, and its value at 2 K is 350 $\mu C/m^2$, 58% smaller than the maximum polarization observed in $TbMnO_3$ after poling in twice higher field (200 $Vmm^{-1}$) [14]. The decrease of the $P_c$ value in $TbMn_{0.98}Fe_{0.02}O_3$ has been reported as a consequence of the destabilization of the magnetic structure due to the changes in magnetic interactions promoted by $Fe^{3+}$ [6]. As the magnetic field increases toward 3 T, $P_c(T)$ shows a maximum, peaking at highest temperatures according to the magnetic field strength, and then shifts to lower temperatures on further magnetic field increase up to 5 T. Interestingly, the value of $P_c$ at fixed temperatures decreases at a faster magnetic field rate for $B > 5.5$ T. The $P_c(T)$ curve changes its shape for at 6 T, fading out with decreasing temperature, in such a way that for 9 T, $P_c \sim 0$ below 13 K. Contrarily to $TbMnO_3$, the magnetic field evolution of $P_c(T)$, with $B \parallel b$ is more gradual in this compound [14].

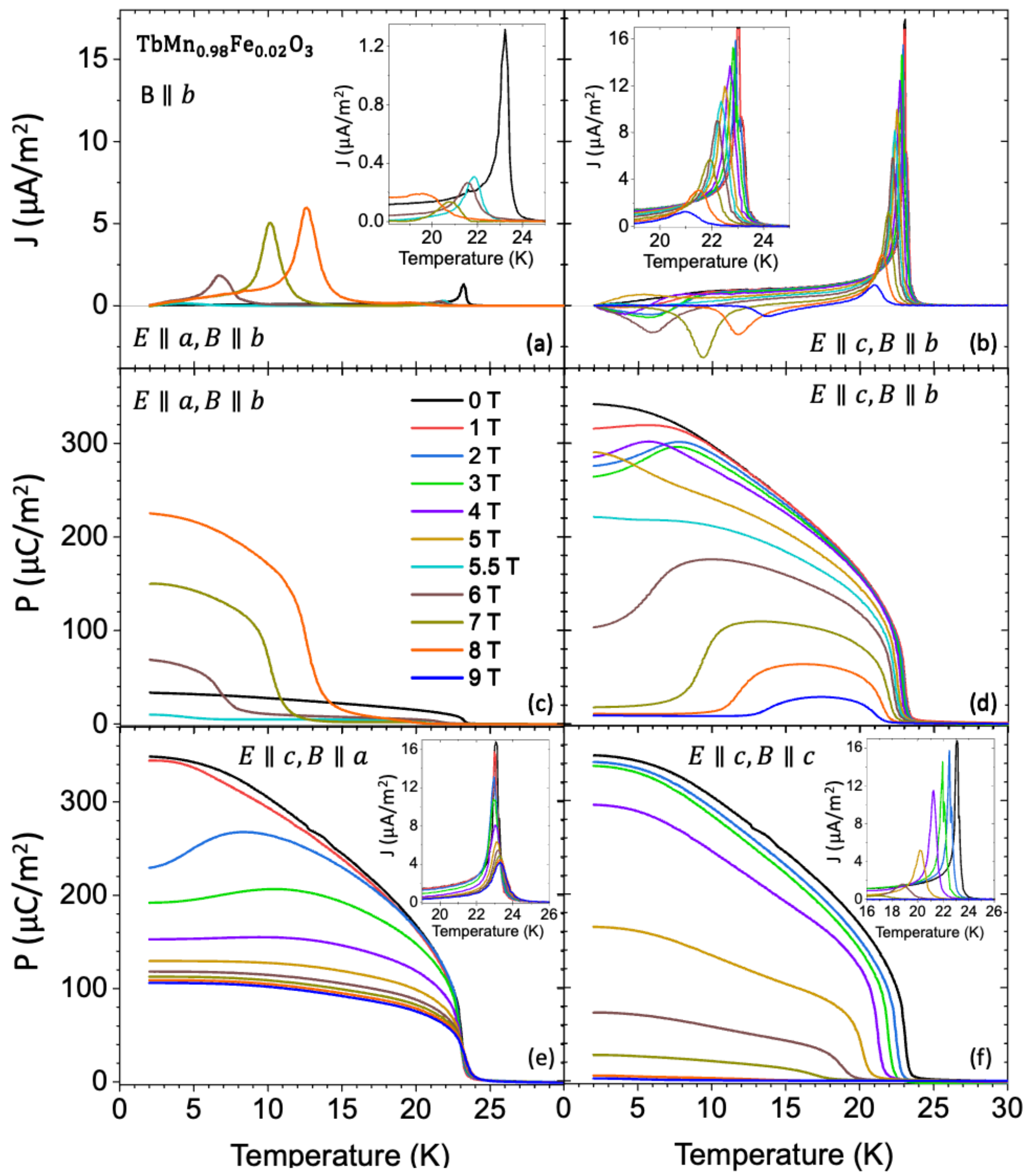

Figure 1. Temperature dependence of the ((a), (b)) pyroelectric current densities and ((c), (d)) electric polarization of $TbMn_{0.98}Fe_{0.02}O_3$, measured along the *a*-and *c*-axes under an applied magnetic field along the *b*-axis. The insets show a magnified view of the temperature profiles of the pyroelectric current density.

Concerning the temperature dependence of the pyroelectric current density recorded along the *a*-axis, $J_a(T)$, a small anomaly at $T_c = 23$ K is already observed for 0 T contrary to pure $TbMnO_3$, where no anomaly was detected [14]. As the magnetic field strength increases, the amplitude of this anomaly decreases and downshifts, following a similar temperature trend as observed for $J_c(T)$ (see inset of Figure 1(a)). Due to the similar temperature behavior, we assign this anomaly to the minor electric polarization projection in the measurement direction, due to a small misorientation in the sample. A second anomaly in $J_a(T)$ emerges at low temperatures with the same sign as the first one. The second anomaly peaks at higher temperatures and its amplitude increases as the magnetic field strength increases. These results evidence the enhancement of the electric polarization along the *a*-axis as the magnetic field increases, as seen in Figure 1 (c). The maximum electric polarization along the *a*-axis, recorded at 2 K and 8 T, is about 70% of the maximum spontaneous polarization value measured at the same temperature along the *c*-axis, in the absence of the applied magnetic field. In $TbMnO_3$, the maximum value of the electric polarization along the *a*-axis is about 67% of the maximum value of the electric polarization from the *c*-axis, after the magnetically induced polarization flop [14].

Figure 1(e) shows the temperature dependence of the electric polarization measured under applied magnetic field along the *a*-axis ($\boldsymbol{B} \parallel a$). The pyroelectric current $J_c(T)$, shown in the inset of Figure 1(e), exhibits just one anomaly, peaking at $T_C = 23$ K, weakly dependent on the field strength. However, the amplitude of this anomaly decreases with increasing field strength; consequently, the electric polarization is a decreasing function of the magnetic field strength, gradually converging to the limit value 110 μC/m$^2$, contrarily to what is observed when the magnetic field is applied along the other two crystallographic axes. At 9 T, the value of $P_c$ at the lowest temperature is 25% of the value obtained in $TbMnO_3$ which, however, was poled at twice higher electric field [14].

Figure 1 (f) depicts the temperature dependence of the electric polarization and pyroelectric current (inset of Figure 1 (f)) measured along the *c*-axis, under different magnetic field strength, applied along the same c-axis. Only one anomaly is observed in $J_c(T)$ for all the values of the field strength; as consequence, at a fixed magnetic field, the electric polarization increases monotonously on cooling below $T_C$. Moreover, the $J_c(T)$ anomaly peaks at $T_c = 23$ K for 0 T, monotonously decreasing the peaking temperature as the magnetic field increases, reaching 16 K for 9 T (see inset of Figure 1 (f)). The amplitude of this anomaly also decreases and eventually disappears for $B > 9$ T, meaning that the electric polarization is a decreasing function of the applied magnetic field along c-axis, reaching negligible values for 9 T. The latter result contrasts with the temperature/magnetic field dependence of the electric polarization of $TbMnO_3$ measured in the same configuration but poled at twice higher electric field [14]. In fact, for $B < 6$ T, $P_c(T)$ is weakly dependent on the magnetic field, increasing with decreasing temperature, although a change of slope of $P_c(T)$ is observed just below 10 K [14]. However, as the magnetic field increases further from 6 T, the $P_c(T)$ curve profile changes and, for 8 and 9 T, the electric polarization takes non-negligible values between $T_C$ and 15 K [14].

### 3.2. Neutron diffraction

In this section, we will address the temperature and magnetic field dependencies of the measured nuclear and magnetic diffraction peaks. Figure 2 shows representative neutron diffraction patterns, recorded at fixed temperatures in the 5 – 50 K range, and fixed magnetic field strength (0 – 8 T range), applied along the *b*-axis. The results recorded for other magnetic field strengths and same temperatures, are shown in Figure S1 of Supplemental Information [16]. Only the upper half part of the detectors panel is presented, with the main observed diffraction peaks indexed. Experimental restrictions, due to the configuration of the magnets, prevented us from collecting enough diffraction peaks to determine the real space spin structure.

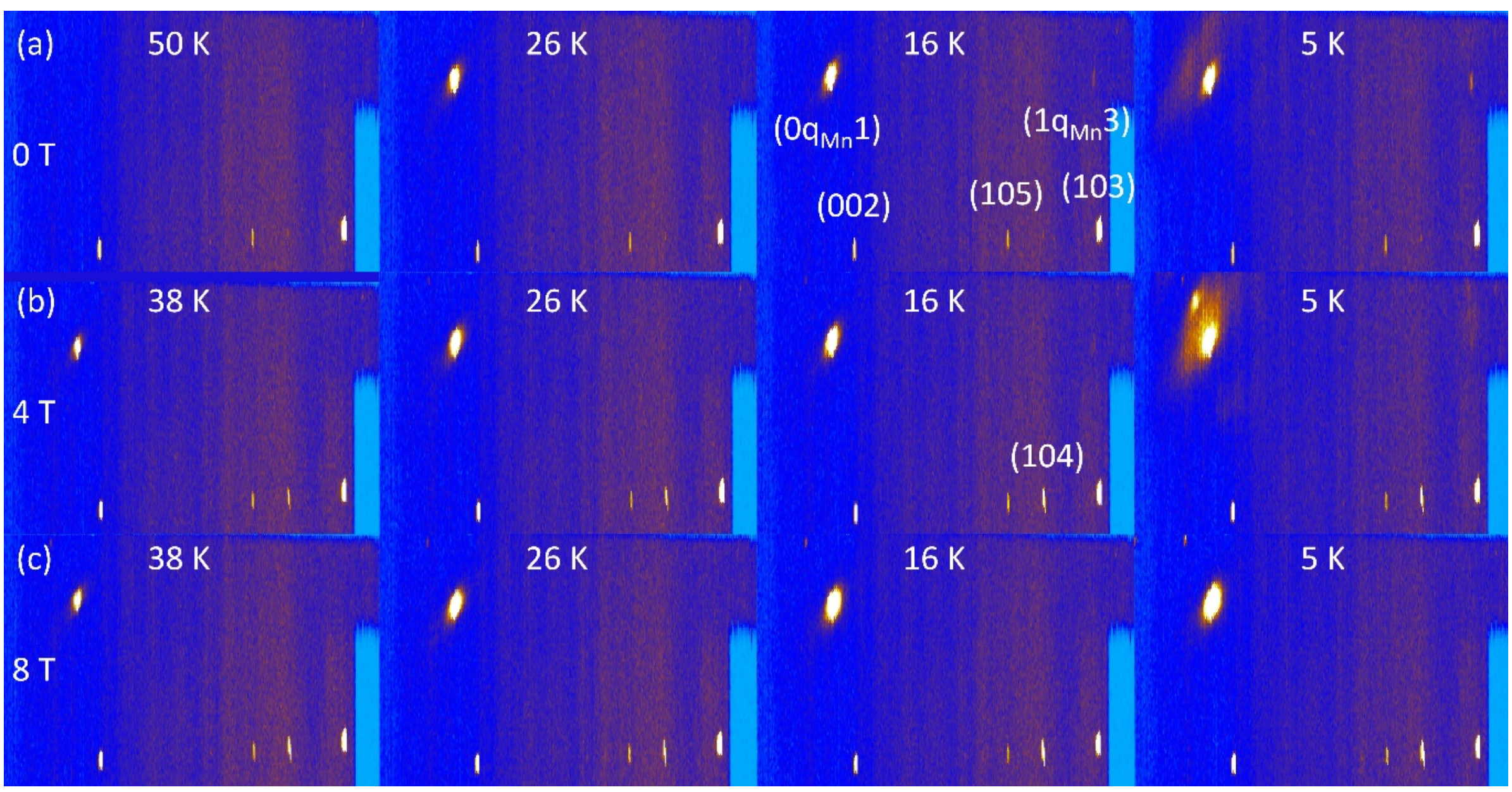

Figure 2. Diffraction patterns of $TbMn_{0.98}Fe_{0.02}O_3$, recorded at different fixed temperatures and magnetic fields of 0 T (a), 4 T (b) and 8 T (c), applied along the *b*-axis.

Above $T_N$ = 39 K, only nuclear peaks are observed and are indexed to reflections from the (0 0 2), (1 0 5) and (1 0 3) crystallographic planes. From the analysis of these peaks, the temperature and magnetic field dependence of the *a* and *c* lattice parameters was determined, as shown in Figure S2 of Supplemental Information for 0 and 8 T, respectively [16]. The lattice parameters slightly decrease with decreasing temperature in the 5 – 50 K range and their values are magnetic field independent. These results evidence for the negligible influence of both the magnetic phase transitions and the applied magnetic field on the lattice parameters, within the sensitivity of the diffraction technique. However, this does not mean that the magnetic phase transitions do not involve atomic motions, because the inversion center must be broken to the electric polarization appear. Recently, spin-phonon coupling was ascertained from a temperature dependent Raman scattering study in $TbMn_{0.98}Fe_{0.02}O_3$, where small anomalies in the temperature dependence of the wavenumber of some phonons assigned to Tb-oscillations and symmetrical stretching mode of the oxygen octahedra were detected [3]. However, the magnetoelastic coupling is rather small in this compound to be observed through neutron diffraction techniques.

As the temperature lowers from $T_N$ = 39 K, well-defined magnetic diffraction peaks appear along with the nuclear diffraction peaks (see Figure 2). This is the case of the magnetic superlattice peak attributed to the $Mn^{3+}$ moment ordering with propagation wavevector (0 $q_{Mn}$ 1) ($q_{Mn}$ is expressed in reciprocal lattice units, r.l.u.), which appears just below $T_N$, as it was also reported for $TbMnO_3$ [9,10,12]. This peak is observed in the 0 – 8 T magnetic field strength range. On further cooling, a second magnetic superlattice peak appears, but only below $T_C$ = 23 K. Following Kajimoto *et al*, we assign this peak to the *G*-type propagation wavevector (1 $q_{Mn}$ 3) of the $Mn^{3+}$ moment ordering structure; the intensity of the (1 $q_{Mn}$ 3) magnetic superlattice peak strongly increases below $T_C$ of $TbMnO_3$, being negligibly small above this temperature [10]. Due to the limited *q*-space covered in this experiment, we could not observe the $Tb^{3+}$ magnetic superlattice peaks. Still, the $Tb^{3+}$ spin ordering is ascertained from the appearance of other magnetic superlattice peaks and a strong diffuse scattering around the (0 $q_{Mn}$ 1) peak. Under applied magnetic field, we observe an additional Bragg peak indexed to (1 0 4) and assigned to a (weak) ferromagnetic spin ordering, likely associated with a spin canting induced by the applied field. In the following, we shall address the temperature and magnetic field dependencies of the magnetic diffraction peaks.

Figure 3(a) shows the temperature dependence of the intensity of the magnetic superlattice peak (0 $q_{Mn}$ 1), recorded at 0, 4, 6 and 8 T, as representative examples. On cooling below $T_N$ = 39 K, the intensity of this reflection increases, and is enhanced by the applied magnetic field, the effect being more pronounced in the temperature range of the ferroelectric phase. A detailed analysis of the temperature dependence of the intensity of the magnetic superlattice peak (0 $q_{Mn}$ 1), reveals a small magnetic field dependent kink at $T_C$, which, for the case of $B$ = 0 T as representative example, can be clearly observed in the inset of Figure 3(a), where the solid line describes the temperature-dependent intensity, $I(T)$, in the $T_C < T < T_N$ range, extrapolated for $T < T_C$. The small but clear deviation of the experimental intensity on cooling below $T_C$ was also reported for $TbMnO_3$ [9,10], and it gives clear evidence for the interplay between magnetism and ferroelectricity in these compounds. On further cooling, the intensity of the reflection exhibits different temperature behavior, depending on the magnetic field strength: while for 0 T and 4 T, a maximum is ascertained at 10 K, a sudden increase of the intensity at 10 K and at 14 K is observed for 6 T and 8 T, respectively. The temperature dependence of the modulation wave number $q_{Mn}$, determined from the magnetic superlattice peak (0 $q_{Mn}$ 1) position, recorded for different field strengths, is presented in Figure 3(b). In the absence of an applied magnetic field and at 38 K, $q_{Mn}$ = 0.278, which is about 18% smaller than $q_{Mn}$ = 0.295 reported for $TbMnO_3$ just below $T_N$, which evidences for a longer magnetic modulation wavelength in $TbMn_{0.98}Fe_{0.02}O_3$ [9,10,17]. In the whole temperature range here explored, $q_{Mn}$ is a non-monotonous function of temperature, although the temperature dependence is a function of the magnetic field. In the $T_C < T < T_N$ interval, $q_{Mn}$ reaches a minimum value at temperatures approaching $T_C$ as the magnetic field increases (32 K for 0 T and 24 K for 8 T). The relative variation of $q_{Mn}$ in the $T_C < T < T_N$ range is 0.7%, a much smaller value when compared with $TbMnO_3$, for which the total variation in the sinusoidal incommensurate phase is 3.7% [17].

In the absence of the applied magnetic field, $q_{Mn}$(T) of $TbMnO_3$ decreases on cooling and remains almost temperature independent below $T_C$, reaching values 0.276 in References [9,10] and 0.28 in Reference [17]. In $TbMn_{0.98}Fe_{0.02}O_3$, $q_{Mn}$(T) slightly increases below 35 K and finally saturates near 0.279. Therefore, the temperature dependence is different between $T_N$ and $T_C$, although the modulation wavevector in the ferroelectric phase takes similar value, within measurement accuracy, in the pure and iron substituted samples.

The most significant changes induced by the magnetic field concerning the temperature dependence of $q_{Mn}$ are observed below $T_C$. For $B \leq 5$ T, $q_{Mn}$(T) is a slowly varying temperature function on cooling in such a way that an almost constant plateau is reached, while for $B \geq 6$ T, $q_{Mn}$ is strongly temperature dependent, increasing as temperature decreases. At 8 T, the relative variation of $q_{Mn}$ between $T_C$ and the base temperature is 0.9 %. The magnetic field threshold (B ~5.5 T) is the one for which the polarization flop is observed (see Figure 1 (c) and Figure 1 (d)). The increasing of $q_{Mn}$(T) on cooling from $T_C$ contrast with the observations in $TbMnO_3$, for which at 5.5 T a decrease on cooling is observed [12].

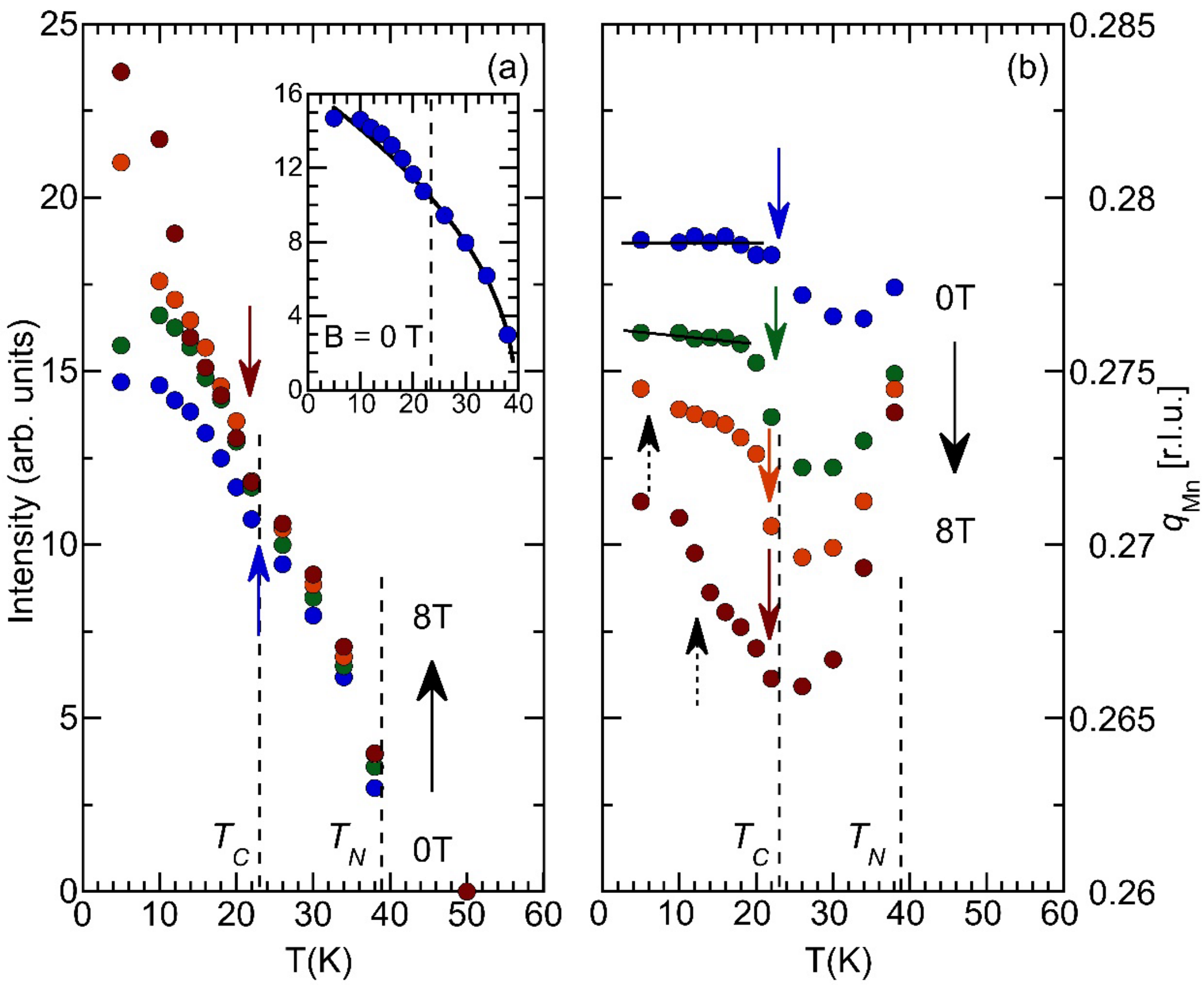

Figure 3. Temperature dependencies of the (a) intensity and (b) $q_{Mn}$ of the magnetic superlattice peak (0 $q_{Mn}$ 1), recorded at 0, 4, 6 and 8 T, as representative examples. The vertical dashed lines mark the $Mn^{3+}$ magnetic ordering temperatures at 0 T. The colored arrows point at the critical temperature $T_C$ where the pyroelectric current density curve exhibits maximum value, while the black dashed arrows in (b) point the temperatures where anomalous increasing of $q_{Mn}$ is ascertained. Inset: temperature dependence of the magnetic superlattice peak measured at B = 0 T. The solid line represents the temperature trend of the peak intensity in the $T_C < T < T_N$ range, extrapolated to $T < T_C$.

These results give clear evidence for the interplay between the mechanisms underlying the stabilization of the ferroelectricity and the modulation of the $Mn^{3+}$ spin structure. However, the main difference here observed relatively to $TbMnO_3$ concerns the effect of the applied magnetic field, having a stronger effect on the modulation wave vector in the Fe-substituted compound in the whole temperature range, while in $TbMnO_3$ the effect is more pronounced in the cycloidal antiferromagnetic phase [12].

In the absence of an applied magnetic field, the (1 0 4) magnetic peak is not observed in the whole temperature range here explored. Under a 1 T strength magnetic field, this reflection appears just below $T_N$ and its intensity increases with decreasing temperature and increasing magnetic field strength, respectively, as it can be seen in Figure 4 (a). The appearance of this magnetic peak clearly points out for the stabilization of a ferromagnetic component with increasing the magnetic field. Concomitantly, the intensity of the *G*-type $Mn^{3+}$ moment ordering magnetic superlattice peak (1 $q_{Mn}$ 3), shown in Figure 4 (b), only observed below $T_C$ but already at B = 0 T, decreases as the magnetic field increases.

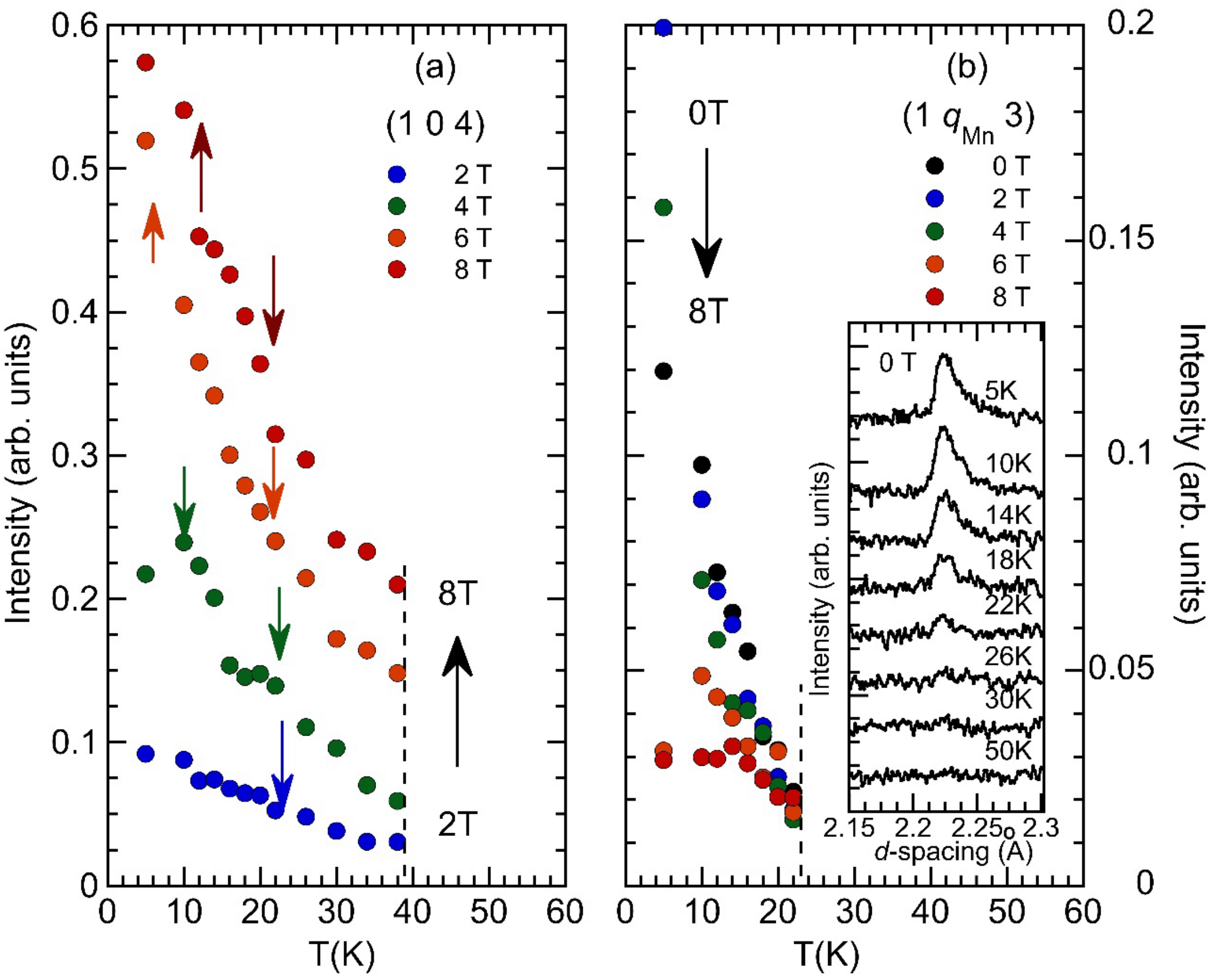

Figure 4. Temperature dependence of (a) the intensity of the (1 0 4) magnetic diffraction peak and (b) the intensity of the magnetic superlattice peak (1 $q_{Mn}$ 3) position for $TbMn_{0.98}Fe_{0.02}O_3$, recorded at different fixed magnetic field strength. The colored arrows point at the critical temperature where the pyroelectric current density curve exhibits maximum value. Inset: neutron diffraction pattern recorded at different fixed temperatures and in the absence of applied field, showing the magnetic superlattice peak (1 $q_{Mn}$ 3).

The strong changes induced by the applied magnetic field in the neutron diffraction patterns depicted in Figure 2, point for magnetic phase transformations below the $Tb^{3+}$ ordering temperature. The diffuse scattering signal, observed around the (0 $q_{Mn}$ 1) magnetic superlattice peak, which prevails up to 4 T, is ascribed to the quite complex and of short-range (local) ordering of $Tb^{3+}$ spins [9,10]. On increasing the magnetic field strength, another magnetic satellite peak starts to be observed, emerging from the diffused scattering signal, as it is shown in Figure 5 for T = 5 K. This peak is indexed to the (0 0.38 1) associated with the modulation of the $Tb^{3+}$ spin structure induced by the applied magnetic field. The intensity and width of this magnetic satellite peak is strongly dependent on the applied magnetic field, and it disappears above 5 T. This result evidences a magnetic field-induced transition of the $Tb^{3+}$ spins, with critical field strength of ~5 T, which deserve to be further studied.

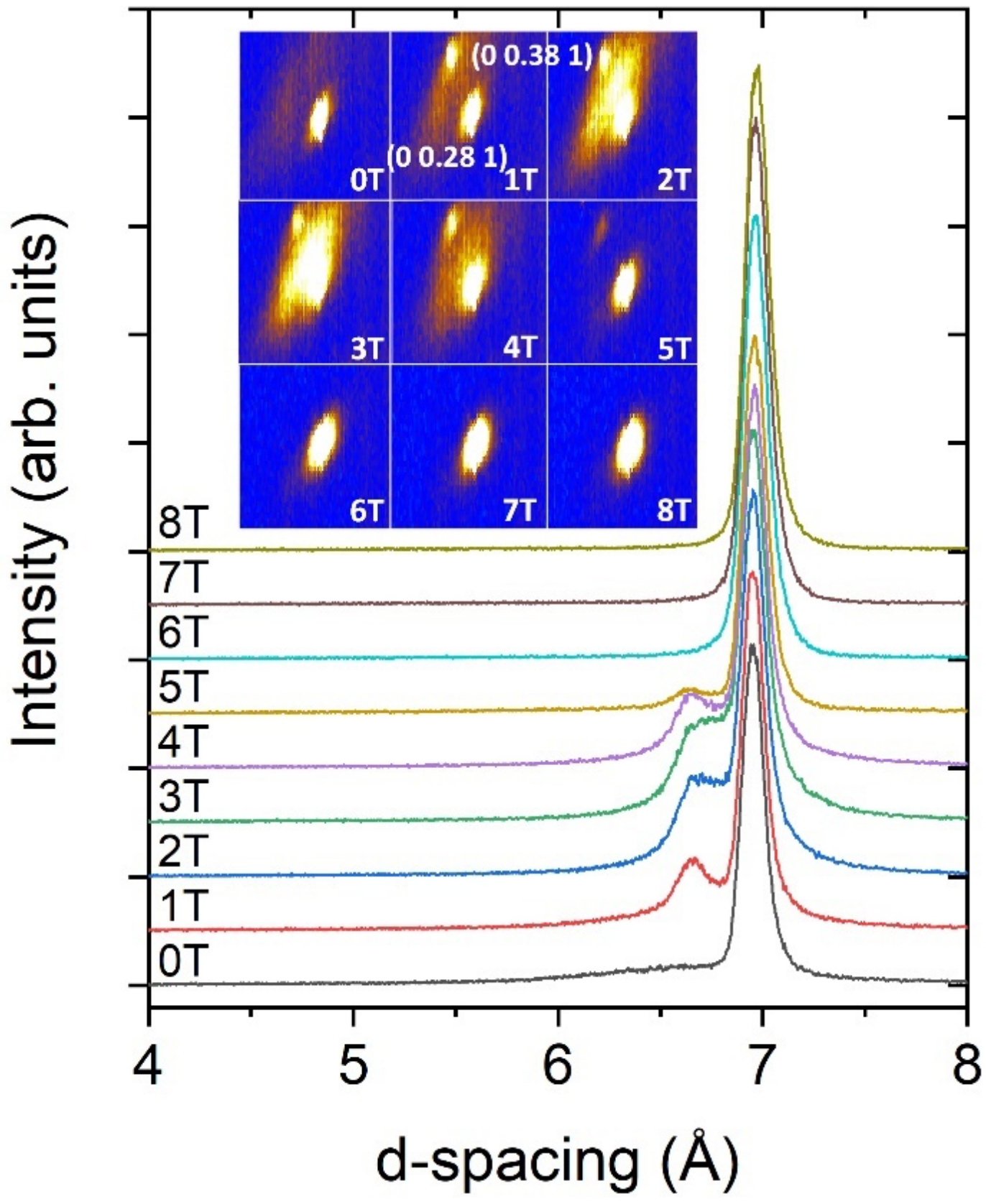

Figure 5. Neutron diffraction patterns of $TbMn_{0.98}Fe_{0.02}O_3$ recorded at 5 K under several applied magnetic fields along the *b*-axis. The inset shows a detailed view of the neutron diffraction patterns around the (0 $q_{Mn}$ 1) peak.

## 4. Discussion

In this section, we will discuss and correlate the results obtained from polarization and neutron diffraction measurements under magnetic field to unravel the microscopic origin of the magnetically induced ferroelectricity in this compound and its magnetic field dependence. We will address the interplay between magnetism and polar properties, highlighted in Figure 6, which shows the temperature dependence of the pyroelectric current density and the intensity of the magnetic superlattice peak (0 $q_{Mn}$ 1), for 0, 4, 6 and 8 T magnetic field strength, applied along the *b*-axis, respectively.

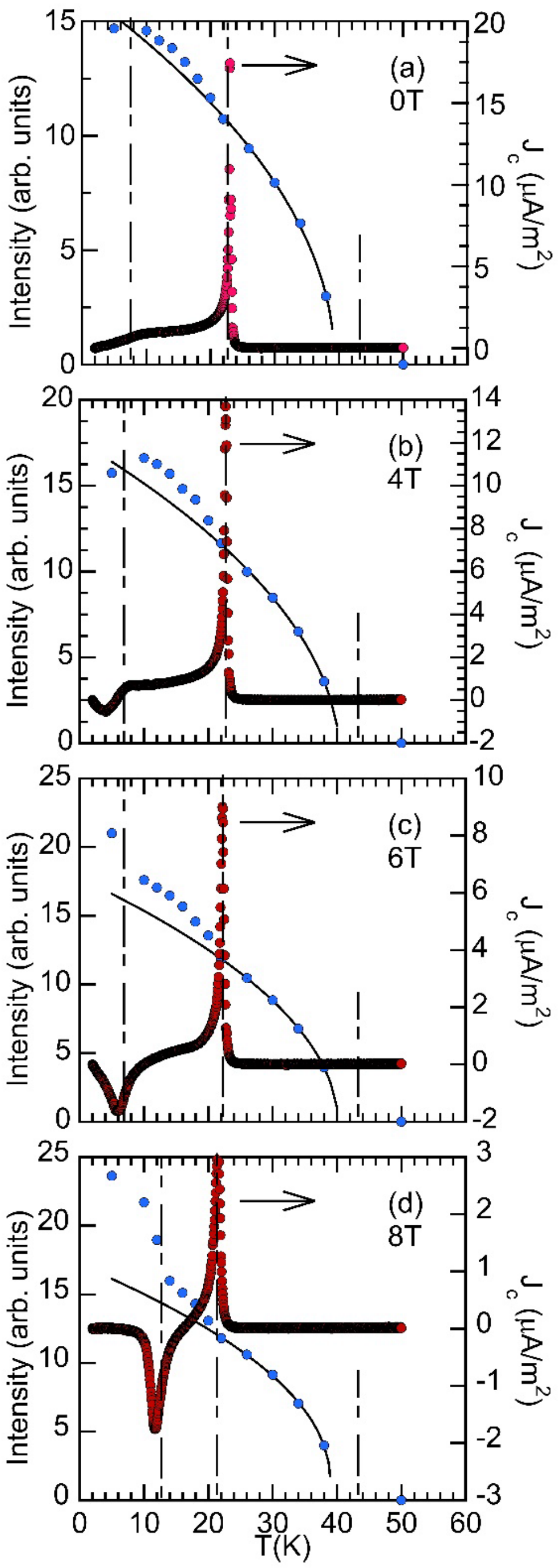

Figure 6. Temperature dependence of the magnetic superlattice peak (0 $q_{\mathrm{Mn}}$ 1) intensity (left) and the pyroelectric current density (right), measured along the *c*-axis in different magnetic fields applied along *b*-axis. The solid line represents the temperature trend of the peak intensity in the $T_{\mathrm{C}} < \mathrm{T} < T_{\mathrm{N}}$ range, extrapolated to $\mathrm{T} < T_{\mathrm{C}}$.

The paramagnetic to incommensurate sinusoidal antiferromagnetic phase transition at $T_{\mathrm{N}}$ is not revealed by any anomaly in the pyroelectric current density, as expected if the inverse Dzyaloshinskii-Moriya mechanism is assumed only below $T_{\mathrm{C}}$, like in $TbMnO_3$ [18,19]. The magnetically induced ferroelectric phase transition is marked by the anomalous temperature dependence of the (0 $q_{Mn}$ 1) magnetic superlattice peak intensity at $T_C$. At zero magnetic field, the small anomaly in $J_c$(T) occurring at 8 K, which is likely associated with the $Tb^{3+}$ spin ordering. The correlation between the anomalous temperature dependence of both $J_c(T)$ and the magnetic superlattice peak (0 $q_{Mn}$ 1) intensity holds even for applied fields, but following different behavior: at 4 T, $J_c(T)$ exhibits a downward peak at 4 K and the magnetic superlattice peak intensity reaches a maximum value at 10 K; for 6 and 8 T, the downward peak of $J_c(T)$, whose amplitude increases with the field strength, is accompanied by a sudden increase of the slope of the temperature dependence of the magnetic superlattice peak intensity. The spin cycloid is a polar magnetic structure, generating electric polarization in the spin-plane, perpendicularly to the propagation vector ($\boldsymbol{P} \sim \boldsymbol{k} \times (\boldsymbol{S}_i \times \boldsymbol{S}_j)$) [18,19]. In the *bc*-cycloid phase, the electric polarization is expected to be along the *c*-axis, while in the *ab*-cycloid phase, to be along the *a*-axis. During the cycloid plane rotation from the *bc*- to the *ab*-plane, it is expected an increase of intensity of the (0 $q_{Mn}$ 1) magnetic superlattice peak, simply because in the experimental geometry here described, neutrons probe only the component of magnetic moment perpendicular to the scattering vector, being this component larger when the spins are confined within the *ab*-plane. The results here presented evidence for a strong interplay between the modulated magnetic $Mn^{3+}$ spin structure and the stabilization of the electric polarization, as expected in type-II multiferroics.

The interplay between the magnetic structure, electric polarization and magnetoelectric coupling is better ascertained by comparative plots at fixed temperatures. For this purpose, we represented in Figure 7 the magnetic field dependence of $P_c(T)$, measured with applied magnetic field along the three crystallographic directions, the modulation wave vector $q_{Mn}$, determined from the magnetic superlattice peak (0 $q_{Mn}$ 1) position, and, in inset, the intensity of the magnetic superlattice (1 $q_{Mn}$ 3) (left axis) and of the ferromagnetic (right axis) peaks. Here, we chose to analyze these quantities as a function of the magnetic field strength at 18 K (for which the *a*-axis component of the polarization is not observed up to 8 T), 12 K (for which the polarization rotates with applied magnetic field), and 5 K (to study the $Tb^{3+}$-spin ordering effect), which represent the overall trend of the aforementioned quantities.

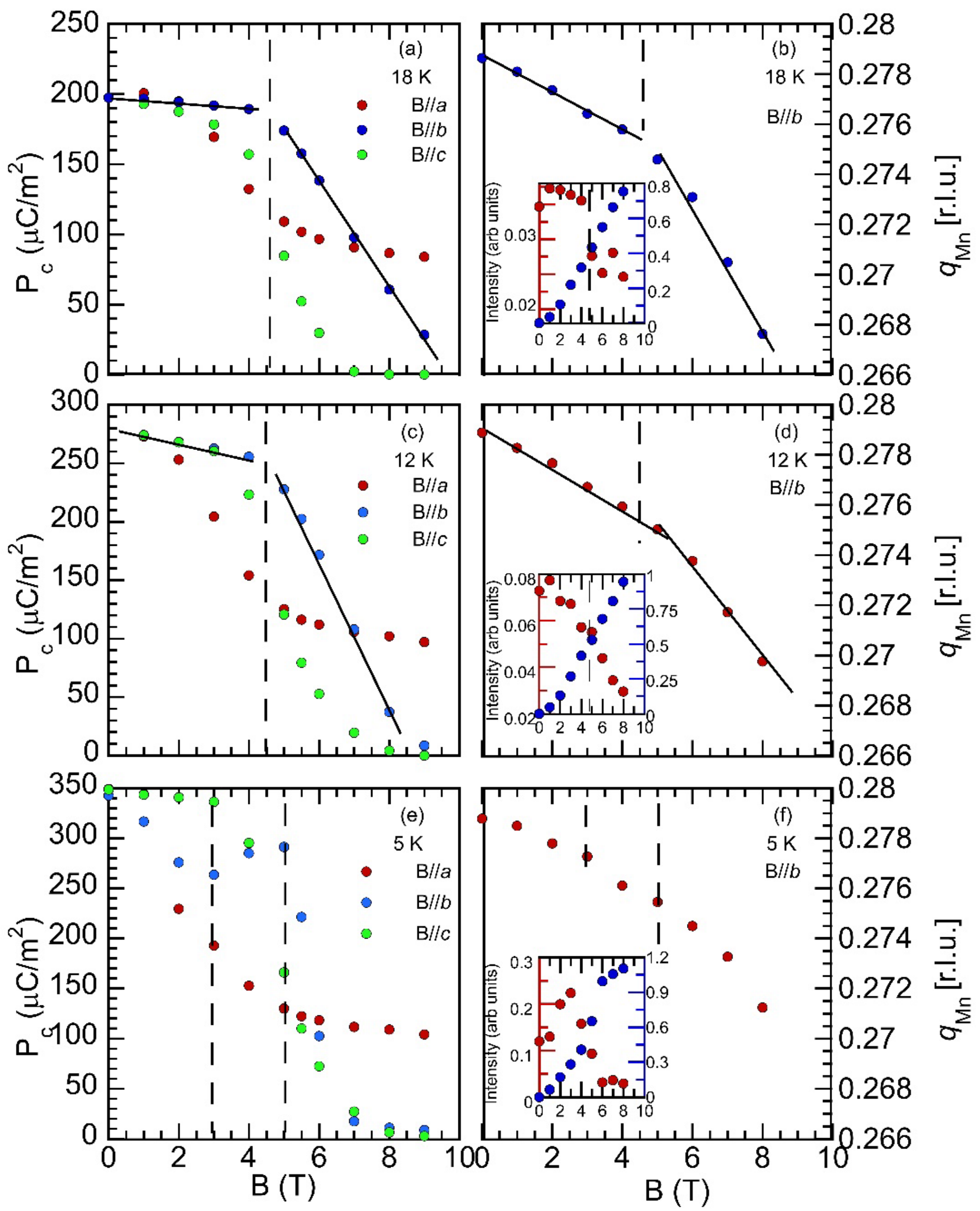

Figure 7. Magnetic field dependence of the electric polarization $\boldsymbol{P}_c$ measured along the $c$-axis in external magnetic field applied along the $a$-, $b$- and $c$-axes, and the modulation wave vector $q_{Mn}$, calculated from the (0 $q_{Mn}$ 1) peak position, measured at 18 K (a, b), 12 K (c, d) and 5 K (e, f). Insets: magnetic field dependence of the intensity of the (1 $q_{Mn}$ 3) magnetic superlattice peak (left red) and the (1 0 4) magnetic peak (right blue).

For the three chosen representative temperatures, $P_c(B)$ exhibits different magnetic field dependences according to the applied magnetic field direction, mirroring the anisotropic nature of the magnetoelectric coupling. In the 8 to 22 K range, $P_c(B)$ is a linear function of B || $b$ below and above 4.5 T, although with different slopes; it is a linear function of $\boldsymbol{B}$ || $a$ only for $B_a < 4$

T; and has a non-linear dependence with $\boldsymbol{B} \parallel c$, eventually disappearing above a certain value of magnetic field which depends on temperature.

From the slopes of the linear relation between $P_c$ and $\boldsymbol{B} \parallel b$ strength ($B_b$), we have estimated the effective magnetoelectric coefficient $\alpha_{32} = dP_c/dB_b$ and its temperature dependence is presented in Figure 8, below and above 4 T.

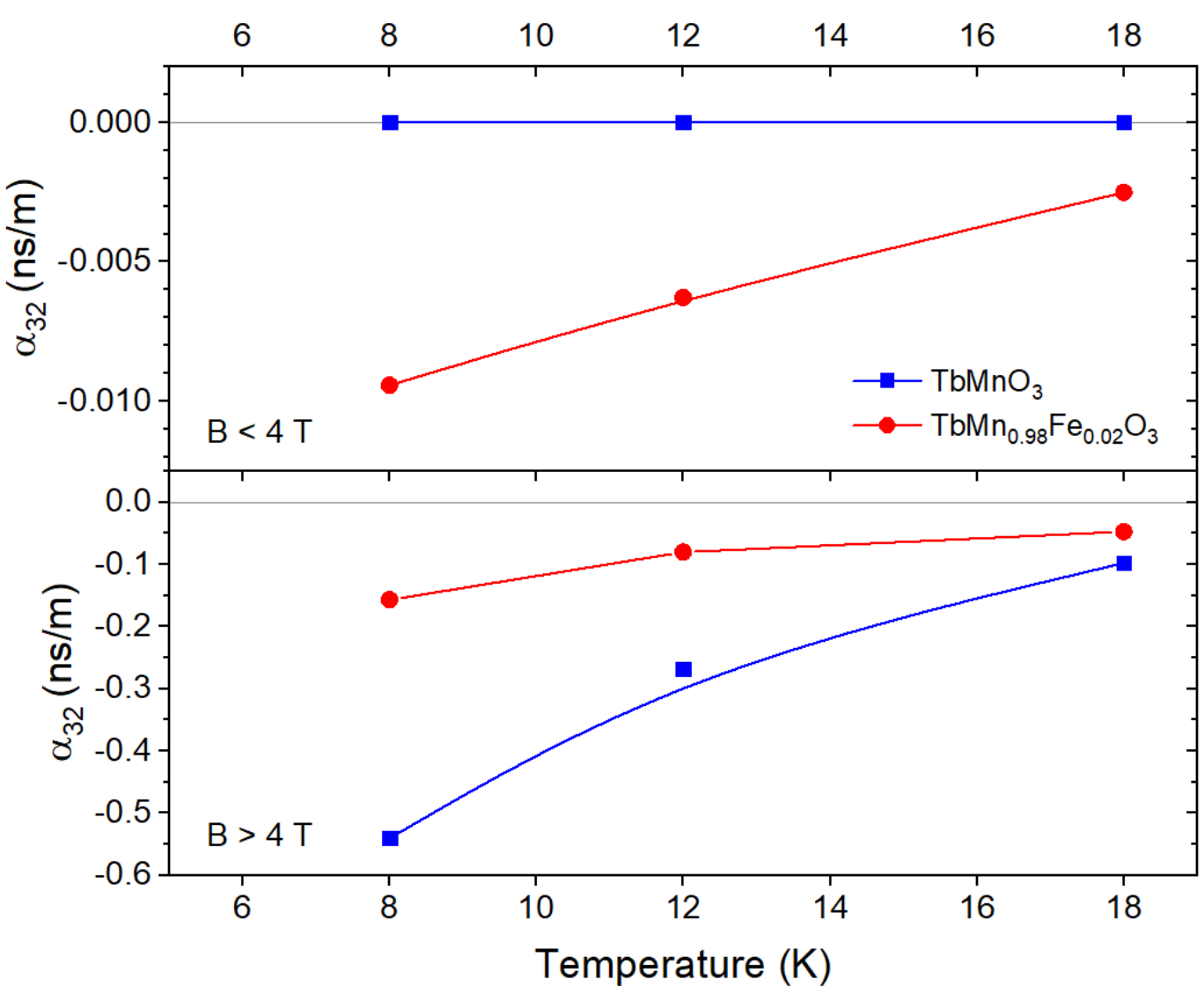

Figure 8. Temperature dependence of the effective magnetoelectric coefficient $\alpha_{32}$ of $TbMnO_3$ and $TbMn_{0.98}Fe_{0.02}O_3$, for (a) $B_b < 4$ T and (b) $B_b > 4$ T. The solid lines are guide for the eyes. Data concerning $TbMnO_3$ was taken from Ref. [14].

The effective magnetoelectric coefficient of $TbMn_{0.98}Fe_{0.02}O_3$ is negative, mirroring the decrease the polarization with the magnetic field increase. The value of the magnetoelectric coefficient is one order of magnitude larger for $B_b > 4$ T. In both magnetic field ranges, the absolute value $\alpha_{32}$ increases monotonously as temperature decreases, giving evidence for the strengthening of the magnetic field dependence of the electric polarization as the temperature/magnetic field decreases/increases down to 8 K.

Comparing with $TbMnO_3$, two different regimes are found, as it is clear from Figure 8. For applied magnetic field strength below 4 T, the effective magnetoelectric coefficient $\alpha_{32} = dP_c/dB_b$ obtained for $TbMnO_3$ is exactly 0, meaning that while the polarization does not rotate towards the $a$-axis, its value is independent of the applied magnetic field. In contrast, $TbMn_{0.98}Fe_{0.02}O_3$ has a finite negative $\alpha_{32}$ coefficient for the same temperature and magnetic field strength. This result proves the role played by the $Fe^{3+}$ magnetism, that even a low level substitution slightly changes the cycloidal ordering, in agreement with the neutron data that show the promotion of a canted structure already at 1 T applied magnetic field. However, above 4 T, the modulus of the effective magnetoelectric coefficient of $TbMnO_3$ becomes higher than the one calculated for $TbMn_{0.98}Fe_{0.02}O_3$, due to a sharper rotation of the cycloid ordering with the applied magnetic field. The difference between both values increasing as temperature decreases, in such a way that at 8 K the effective magnetoelectric coefficient of the pure compound is almost 4 times larger than in the Fe-substituted compound. This demonstrates the strong impact of low-level substitution of $Mn^{3+}$ by $Fe^{3+}$ in the electric polarization dependence on the low applied magnetic field.

Figures 7(b) and 7(d) show the magnetic field dependence of the modulation wave vector $q_{\mathrm{Mn}}$, recorded at 18 K and 12 K, respectively. Two linear regimes in $\boldsymbol{q}_{\mathrm{Mn}}(\boldsymbol{B}_{\mathrm{b}})$ are ascertained below and above 4.5 T, with a slope increasing from the low to high magnetic field ranges. It is worth to stress that the magnetic field dependence of both $P_{\mathrm{c}}(\boldsymbol{B}_{\mathrm{b}})$ and $q_{\mathrm{Mn}}(\boldsymbol{B}_{\mathrm{b}})$ is rather similar, evidencing for a correlation between $P_{\mathrm{c}}$ and the modulated magnetic structure. To discuss the microscopic origin of the magnetically induced ferroelectricity in this compound, we have analyzed the correlation between the electric polarization $P_{\mathrm{c}}$ and the (0 $q_{\mathrm{Mn}}$ 1) magnetic superlattice peak characteristics, as it better mirrors the cycloidal spin structure underlying the electric polarization. Following the grounds of the spin current model, presented by Y. Yamasaki *et al* [20], the electric polarization is predicted to be proportional to the product of $\sin(2\pi q_{Mn})$ and the square of the magnetic structural factor. Unfortunately, the geometric properties of the spin cycloid in this compound could not be determined from our experimental results, and the accurate determination of the magnetic structural factor is not possible. However, we can overcome this difficulty looking at the intensity $I$ of the (0 $q_{\mathrm{Mn}}$ 1) magnetic superlattice peak, which is proportional to the square of the magnetic structural factor. Figure 9(a) shows $P_{\mathrm{c}}$ as a function of $I \times \sin(2\pi q_{Mn})$, measured at different temperatures and 0 T. The linear regime observed in the absence of applied field evidences that the mechanism underlying the magnetoelectric coupling is based on the spin current model applied to a noncolinear spin structure. Interesting is to study the same dependence but at constant temperature and for different applied magnetic field. This study is performed at 18 K where the polarization flip is not observed in the magnetic field range explored in this work. As it can be observed in Figure 9(b), for $B < 4$ T, a linear regime is already found, evidencing that the same microscopic mechanism also prevails in this magnetic field range. However, for $B > 4$ T, a clear deviation of the linear regime is ascertained, enabling us to conclude that the magnetoelectric coupling for $B > 4$ T is not accurately described by the model in Ref. [20]. Actually, the linear dependence of $P_C$ on $I \times \sin(2\pi q_{Mn})$ is broken for magnetic fields above 4 T, as it can be observed in the inset of Figure 9(b). It is worth to stress that $q_{\mathrm{Mn}}$ is quite temperature independent for B < 4 T, while it becomes strongly temperature dependent for higher fields (see Figure 3(b)). The change of magnetic modulation properties at ~ 4 T is undoubtedly in the basis of the change of microscopic mechanisms underlying ferroelectricity. No similar study has been published in $TbMnO_3$, to the best of our knowledge, and therefore this work provides a better understanding of the influence of the applied magnetic field on the microscopic mechanisms that induce electric polarization in this class of multiferroics.

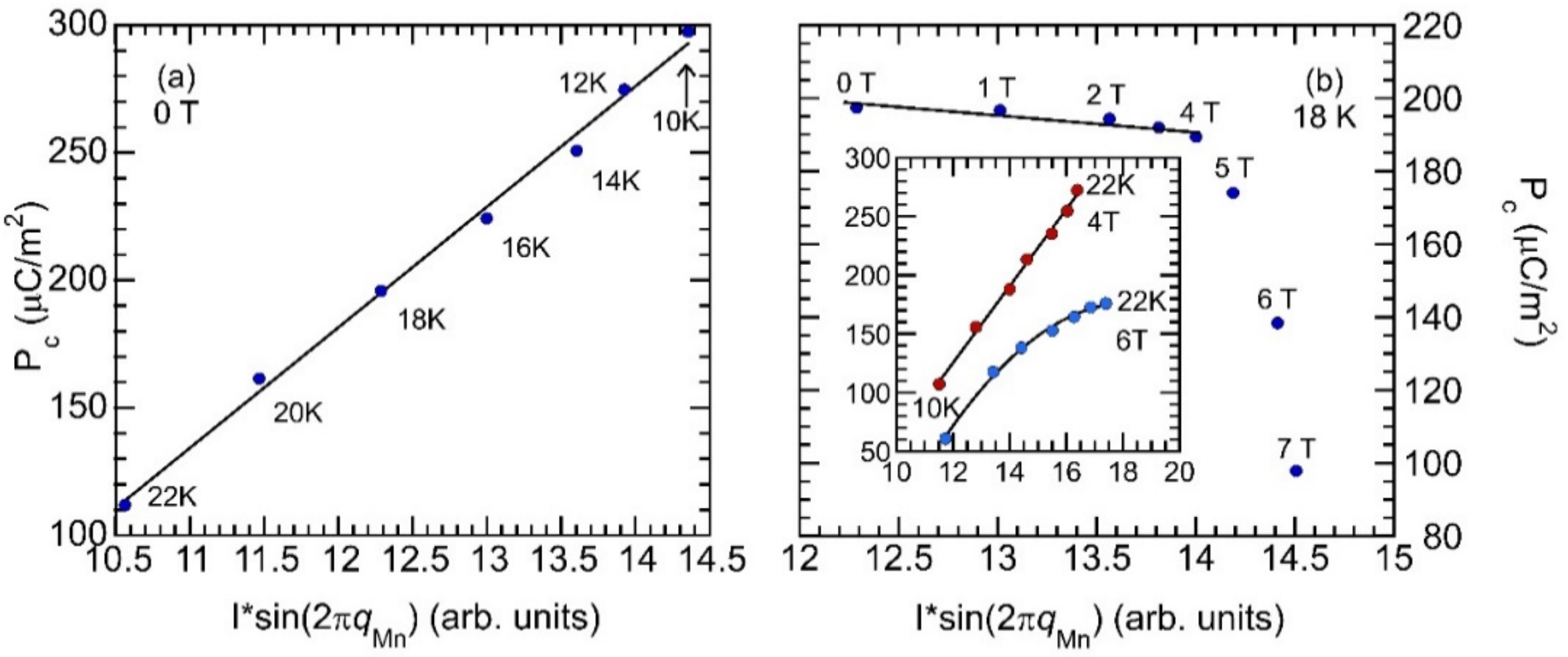

Figure 9. $P$c versus $I \times sin(2\pi q_{Mn})$ recorded at 0 T for different temperatures (a) and at 18 K with magnetic field $\boldsymbol{B}$ || b (b). The intensity $I$ and $q_{Mn}$ values were calculated from the (0 $q_{\mathrm{Mn}}$ 1) magnetic superlattice peak. Inset: $P$c versus $I \times sin(2\pi q_{Mn})$ recorded at 4 T and 6 T, for different temperatures.

Now, we focus on the decrease of $P_c$ of $TbMn_{0.98}Fe_{0.02}O_3$ for $\boldsymbol{B}$ || $b$ at 18 K with increasing magnetic field strength. It is worth to stress that at this temperature and for maximum magnetic field strength here applied, no polarization flop occurs. At this temperature, the intensity of the ferromagnetic reflection is a linear increasing function of $\boldsymbol{B}$ || $b$, and no anomaly is ascertained at ~ 4.5 T (see inset of Figure 7(b)), which gives clear evidence for the increase of the ferromagnetic component induced by the applied magnetic field, likely due to a spin canting. The destabilization of the cycloidal $Mn^{3+}$ spin modulation, along with the concomitant appearance of a ferromagnetic component, explain the decrease of the $\boldsymbol{P}_c$. This result obtained for $TbMn_{0.98}Fe_{0.02}O_3$ contrasts with the one obtained for pure $TbMnO_3$ from which we conclude that the magnetic field promoted spin canting is a consequence of the Fe spins, even in only 2% concentration [14]. Similar magnetic changes are perceived down to 8 K. Future experiments will be performed to confirm this interpretation.

The magnetoelectric coupling at 5 K exhibits different properties. Strictly, $q_{\mathrm{Mn}}$ is no longer a linear function of the applied magnetic field (see Figure 7(f)) and a hint of a downward jump is ascertained at 3.5 T. Interestingly, the magnetic field dependence of the *G*-type magnetic superlattice peak (1 $q_{Mn}$ 3) intensity exhibits a maximum value at 3 T and becomes magnetic field independent above 5 T, while the (1 0 4) magnetic reflection intensity exhibits an s-shaped anomaly at 5 T. Revisiting Figure 7(f), we can assign the aforementioned intensity changes at 5 T to the magnetically induced phase transition. The different behavior described to the one above is a consequence of the $Tb^{3+}$ magnetic ordering on the magnetoelectric coupling at 5 K.

To summarize the results here presented, we propose the magnetoelectric phase diagram of $TbMn_{0.98}Fe_{0.02}O_3$ shown in Figure 10, with magnetic fields along the (a) *a*-, (b) *b*-, and (c) *c*-axes.

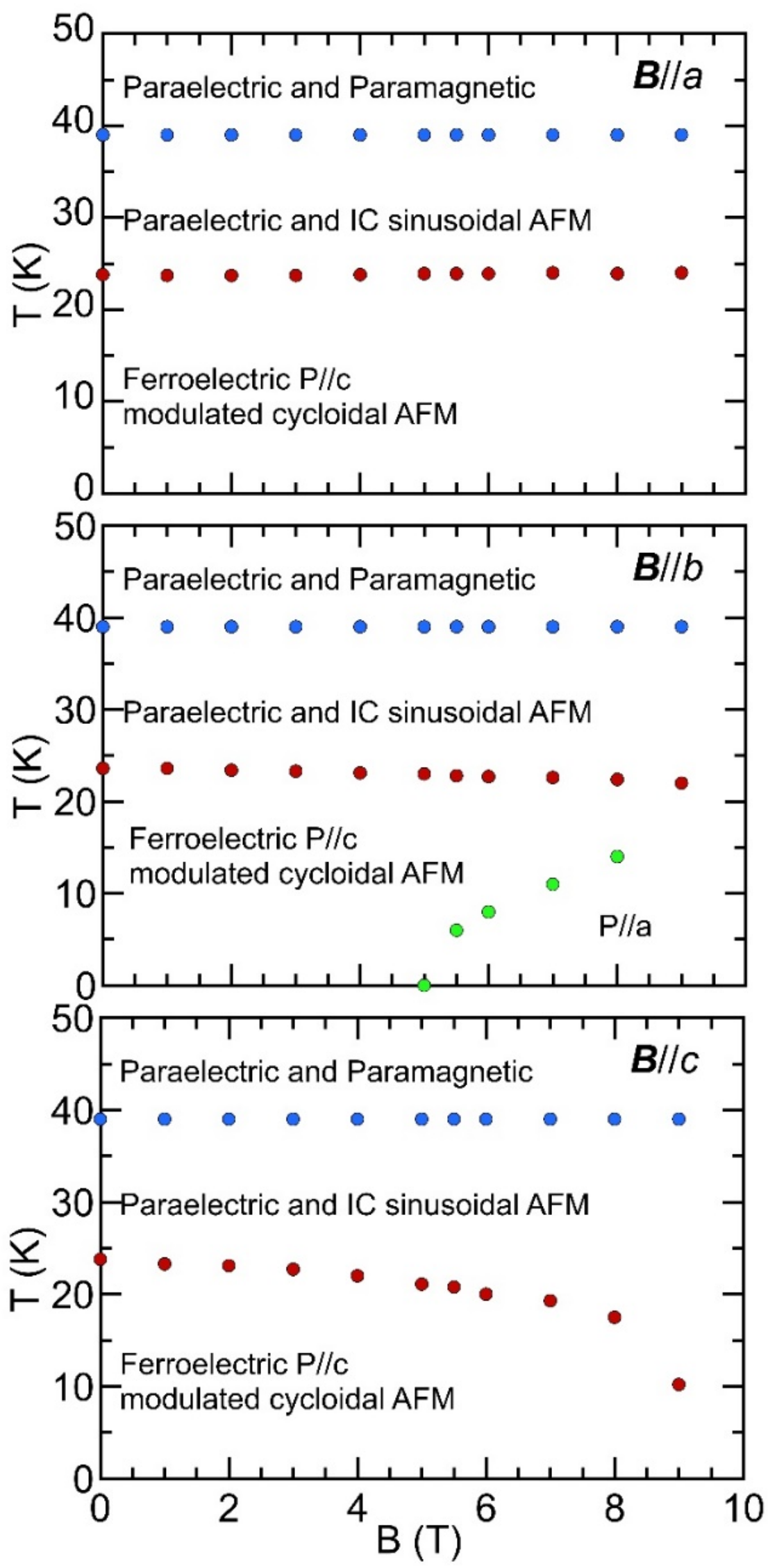

Figure 10. The magnetoelectric phase diagram of $TbMn_{0.98}Fe_{0.02}O_3$, with magnetic fields along the (a) *a*-, (b) *b*-, and (c) *c*-axes. Antiferromagnetic structure has incommensurate sinusoidal modulation between $T_N$ and $T_C$ and cycloidal modulation below $T_C$.

## 5. Conclusions

In this work we report a combined experimental study of the temperature and magnetic field dependence of polarization and magnetic structures in $TbMn_{0.98}Fe_{0.02}O_3$ single crystal. The results show that the low-level of $Mn^{3+}$ substitution for $Fe^{3+}$, although does not change the phase sequence regarding $TbMnO_3$, it causes a higher magnetic field sensitivity of the electric polarization for $B < 4$ T. This is due to the changes on the magnetic modulation of the $Mn^{3+}$ spin structure. Two main consequences arise from the magnetic structure changes: i) the value of the electric polarization is lower than in $TbMnO_3$, and ii) the magnetoelectric coupling coefficient $\alpha_{32}$ at low magnetic fields below 4 T is finite, ranging between -0.005 ns/m to -0.01 ns/m, contrasting with the vanishing value reported for $TbMnO_3$, meaning that while the polarization does not rotate towards the $a$-axis, its value is independent of the applied magnetic field. We attribute this magnetoelectric performance enhancement to the effect of the $Fe^{3+}$ magnetism which alters the magnetic interactions in such extent that the magnetic structure allowing for ferroelectricity, according to Dzyaloshinskii–Moriya interaction, is no longer stable at high magnetic fields. Our experimental work evidences that the low-level substitution of $Mn^{3+}$ by $Fe^{3+}$, though keeping the same phase sequence, induces significant changes in properties and evidences for the rather delicate balance between the competitive magnetic interactions underlying the ground state of $TbMnO_3$.

## Acknowledgments

This work was supported by the VEGA 2/0137/19 project, the Czech Science Foundation (Project No. 21-06802S), Project from the European Union and the Czech Ministry of Education, Youth and Sports (Project TERAFIT-CZ.02.01.01/00/22 008/0004594), PTDC/FIS-MAC/29454/2017, NORTE/01/0145/FEDER/028538, IFIMUP: Norte-070124-FEDER-000070;NECL: NORTE-01-0145- FEDER-022096, UIDB/04968/2020, UID/NAN/50024/2019, PTDC/NAN-MAT/28538/2017 and PTDC/FIS/03564/2022.

## Competing interests

The authors declare no competing interests.

# Supplemental Information

## Strong impact of low-level substitution of Mn by Fe on the magnetoelectric coupling in $TbMnO_3$

A. Maia[1], R. Vilarinho[2], P. Proschek[3], M. Lebeda[1], M. Mihalik jr.[4], M. Mihalik[4], P. Manuel[5], D. D. Khalyavin[5], S. Kamba[1], J. Agostinho Moreira[2]

[1]*Institute of Physics of the Czech Academy of Sciences, Na Slovance 2, 182 00 Prague, Czech Republic*

[2]*IFIMUP, Physics and Astronomy Department, Faculty of Sciences, University of Porto, Porto, Portugal*

[3]*Faculty of Mathematics and Physics, Charles University, Ke Karlovu 5, 121 16 Prague, Czech Republic*

[4]*Institute of Experimental Physics Slovak Academy of Sciences, Watsonova 47, Košice, Slovak Republic*

[5]*ISIS Facility, Rutherford Appleton Laboratory, Harwell Campus, Didcot OX11 0QX, UK.*

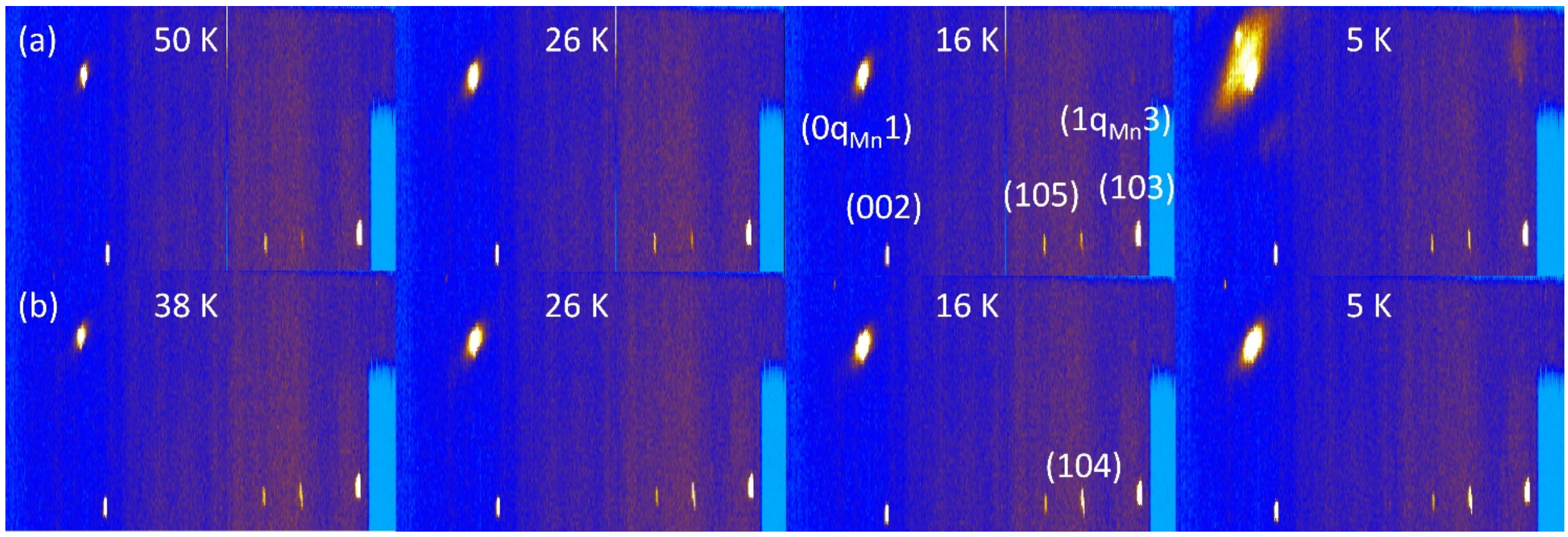

Figure S1. Diffraction patterns of $TbMn_{0.98}Fe_{0.02}O_3$, recorded at different fixed temperatures and magnetic fields of (a) 2 T and (b) 6 T, applied along the *b*-axis.

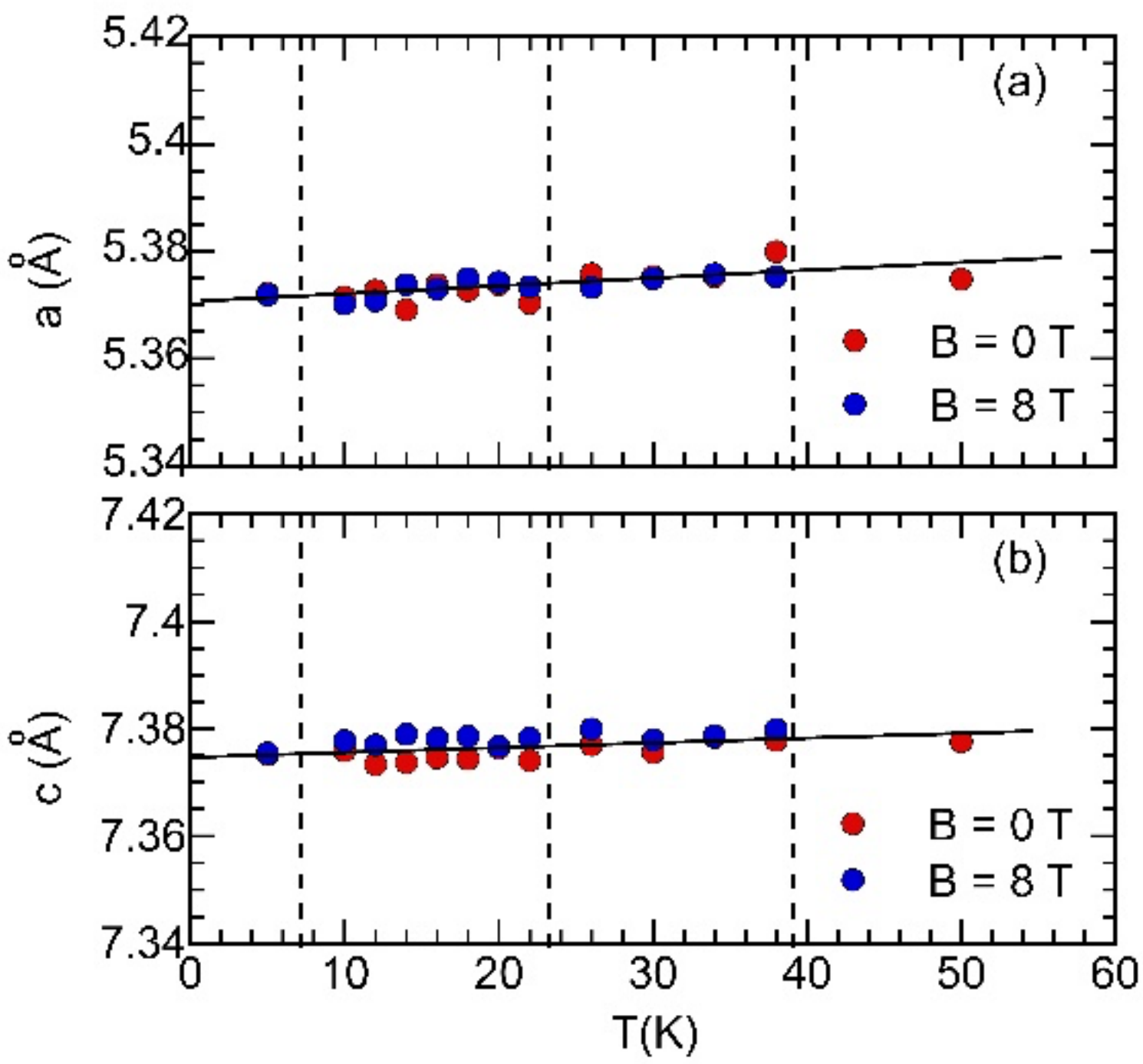

Figure S2. Temperature dependence of the *a* (a) and *c* (b) lattice parameters of $TbMn_{0.98}Fe_{0.02}O_3$, measured at 0 T (red) and 8 T (blue), respectively. The vertical dashed lines mark the critical temperatures and the solid line is a guide for the eyes.

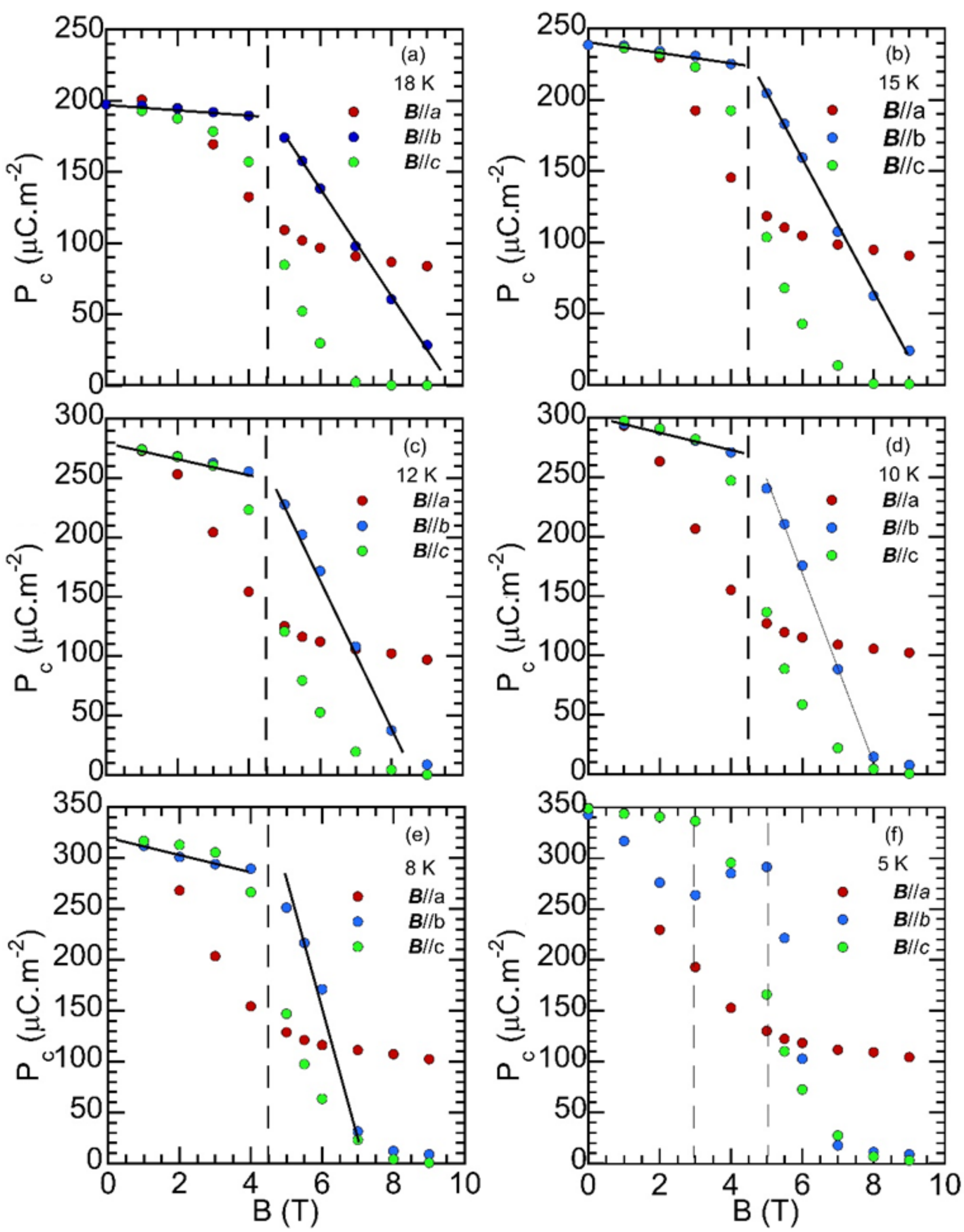

Figure S3. Magnetic field dependence of the electric polarization $\boldsymbol{P}_c$ measured along the *c*-axis in external magnetic field applied along the *a*-, *b*- and *c*-axes.